\documentstyle[twocolumn,epsfig,prl,aps]{revtex}

\long\def\wideabs#1{\twocolumn[\hsize\textwidth\columnwidth\hsize%
\csname @twocolumnfalse\endcsname #1 \vskip1pc]}

\newcommand{\kmmp}{$K^+\rightarrow \mu^+\mu^+\pi^-$ } 
\newcommand{\keep}{$K^+\rightarrow e^+e^+\pi^-$ } 
\newcommand{\kmep}{$K^+\rightarrow\mu^+e^+\pi^-$ } 
\newcommand{\kpem}{$K^+\rightarrow\pi^+e^+\mu^-$ } 
\newcommand{\kefour}{$K^+\rightarrow\pi^+\pi^- e^+ \nu$ }

\newcommand{\kpp}{$K^{+}\rightarrow\pi^{+}\pi^{0}$}

\newcommand{\kpee}{$K^{+}\rightarrow\pi^{+}e^+e^-$}
\newcommand{\kenee}{$K^{+}\rightarrow  e^{+}\nu e^+e^-$}
\newcommand{\kmnee}{$K^{+}\rightarrow\mu^{+}\nu e^+e^-$}
\newcommand{\kpme}{$K^{+}\rightarrow\pi^{+}\mu^+e^-$}
\newcommand{\kpmm}{$K^{+}\rightarrow\pi^{+}\mu^+\mu^-$}

\newcommand{\ktau}{$K^{+}\rightarrow\pi^{+}\pi^+\pi^-$}
\newcommand{\kpitwo}{$K^{+}\rightarrow\pi^{+}\pi^0$}
\newcommand{\pizem}{$\pi^{0}\rightarrow e^+\mu^-$}
\newcommand{\pzem}{$\pi_{e\mu}$}

\newcommand{\efour}{$K_{e4}$}
\newcommand{\pitwo}{$K_{\pi 2}$}
\newcommand{\mmp}{$K_{\mu\mu\pi}$}
\newcommand{\piem}{$K_{\pi e\mu}$}
\newcommand{\mep}{$K_{\mu e\pi}$}
\newcommand{\eep}{$K_{e e\pi}$}
\newcommand{\pmm}{$K_{\pi \mu\mu}$}
\newcommand{\taus}{$K_{\tau}$}

\begin{document}
\draft

\wideabs{
\title{
Search for Lepton Flavor Violation in $K^+$ Decays 
}
\author{
R.~Appel$^{6,3}$, G.~S.~Atoyan$^4$, B.~Bassalleck$^2$,  
D.~R.~Bergman$^6$\cite{DB}, N.~Cheung$^3$, 
S.~Dhawan$^6$,   
H.~Do$^6$, J.~Egger$^5$, S.~Eilerts$^2$\cite{SE},    
H.~Fischer$^2$\cite{HF}, W.~Herold$^5$, 
V.~V.~Issakov$^4$, H.~Kaspar$^5$, D.~E.~Kraus$^3$, 
D.~M.~Lazarus$^1$, 
P.~Lichard$^3$, J.~Lowe$^2$, J.~Lozano$^6$\cite{JL},   
H.~Ma$^1$, 
W.~Majid$^6$\cite{WMa}, W.~Menzel$^7$\cite{WMe},  
S.~Pislak$^{8,6}$, A.~A.~Poblaguev$^4$, P.~Rehak$^1$,  
A.~Sher$^3$ J.~A.~Thompson$^3$, 
P.~Tru\"ol$^{8,6}$, and M.~E.~Zeller$^6$   
}

\address{
$^1$ Brookhaven National Laboratory, Upton, NY 11973, USA\\ 
$^2$Department of Physics and Astronomy, 
University of New Mexico, Albuquerque, NM 87131, USA\\
$^3$ Department of Physics and Astronomy, University of Pittsburgh,
Pittsburgh, PA 15260, USA \\ 
$^4$Institute for Nuclear Research of Russian Academy of Sciences, 
Moscow 117 312, Russia \\
$^5$Paul Scherrer Institut, CH-5232 Villigen, Switzerland\\ 
$^6$ Physics Department, Yale University, New Haven, CT 06511, USA\\
$^7$Physikalisches Institut, Universit\"at Basel, CH-4046 Basel,
Switzerland\\
$^8$ Physik-Institut, Universit\"at Z\"urich, CH-8057 Z\"urich,
Switzerland}
 \date{\today} 
\maketitle

\begin{abstract}
A search for lepton flavor violating decays, \kmmp, \keep, \kpem,
\kmep\, and \pizem, was performed using the data collected in E865 at 
the Brookhaven Alternating
Gradient Synchrotron.  No signal was found in any of the decay modes.
At the 90\% confidence level, the branching ratios are less than 
$3.0\times10^{-9}$, $6.4\times10^{-10}$, $5.2\times10^{-10}$, 
$5.0\times10^{-10}$ and $3.4\times10^{-9}$ respectively. 
\end{abstract}

}

The apparent lepton flavor conservation observed so far in particle
physics is conveniently accommodated in the Standard Model if the
neutrino masses are zero.  Such symmetry can be broken by new
physics at a higher energy scale, such as 
Technicolor or Supersymmetry, 
or by neutrinos having Majorana masses.
Extensive experimental efforts have been devoted to searches for lepton flavor
violating kaon decays, 
$K^0_L\rightarrow \mu^\pm e^\mp$\cite{e871} and 
$K^+\rightarrow \pi^+\mu^+ e^-$\cite{e865pme}. 
In this letter, we report the results of a search for 
\kmmp(\mmp), \keep(\eep), \kmep(\mep)\, and \kpem(\piem).
Unlike \kpme, which only violates lepton flavor conservation, 
these decays also violate  generation number conservation.
In addition, the first three decays violate total lepton number 
conservation. 
\mmp\, and \eep\, can proceed by the same
mechanism as neutrinoless double $\beta$-decays of nuclei if neutrinos 
have Majorana masses. 
Although the first generation is well explored in 
neutrinoless double $\beta$-decays, \mmp\, provides a unique  channel to
 search for effects of Majorana neutrinos in the second generation
\cite{litt}.

The previous searches for 
\eep, \mep\, and \piem\, were performed at CERN 25 years ago
\cite{diamant}.
At the 90\% confidence level (C.L.), the branching ratios were found to be 
$Br$(\keep) $<  1\times10^{-8}$,
$Br$(\kpem) $<  7\times10^{-9}$ and 
$Br$(\kmep) $<  7\times10^{-9}$.
In a reanalysis of data of a 1968 bubble chamber experiment
\cite{chang}, the best limit on \mmp\, was determined to be 
 $Br$(\kmmp) $  <  1.5\times10^{-4}$ at the 90\% C.L.\cite{litt}.

Experiment E865 at the Brookhaven Alternating
Gradient Synchrotron was primarily designed to search for 
\kpme\cite{e865pme}.  Because of its excellent capability in kinematic reconstruction
and particle identification of 
$K^+$ decays to three charged particles, it has been exploited to 
study other decays such as \kpee \cite{e865piee}, \kenee\ and \kmnee. 
In 1997, two special data sets were collected  to study 
\kpmm(\pmm) and \kefour(\efour).  Over 400 \pmm\, events \cite{e865pimm} and
400,000
\efour\, events were observed.  We use the former to search for \mmp, and
the
latter for \eep, \piem\ and  \mep.

\begin{figure}[htb]
\epsfig{figure=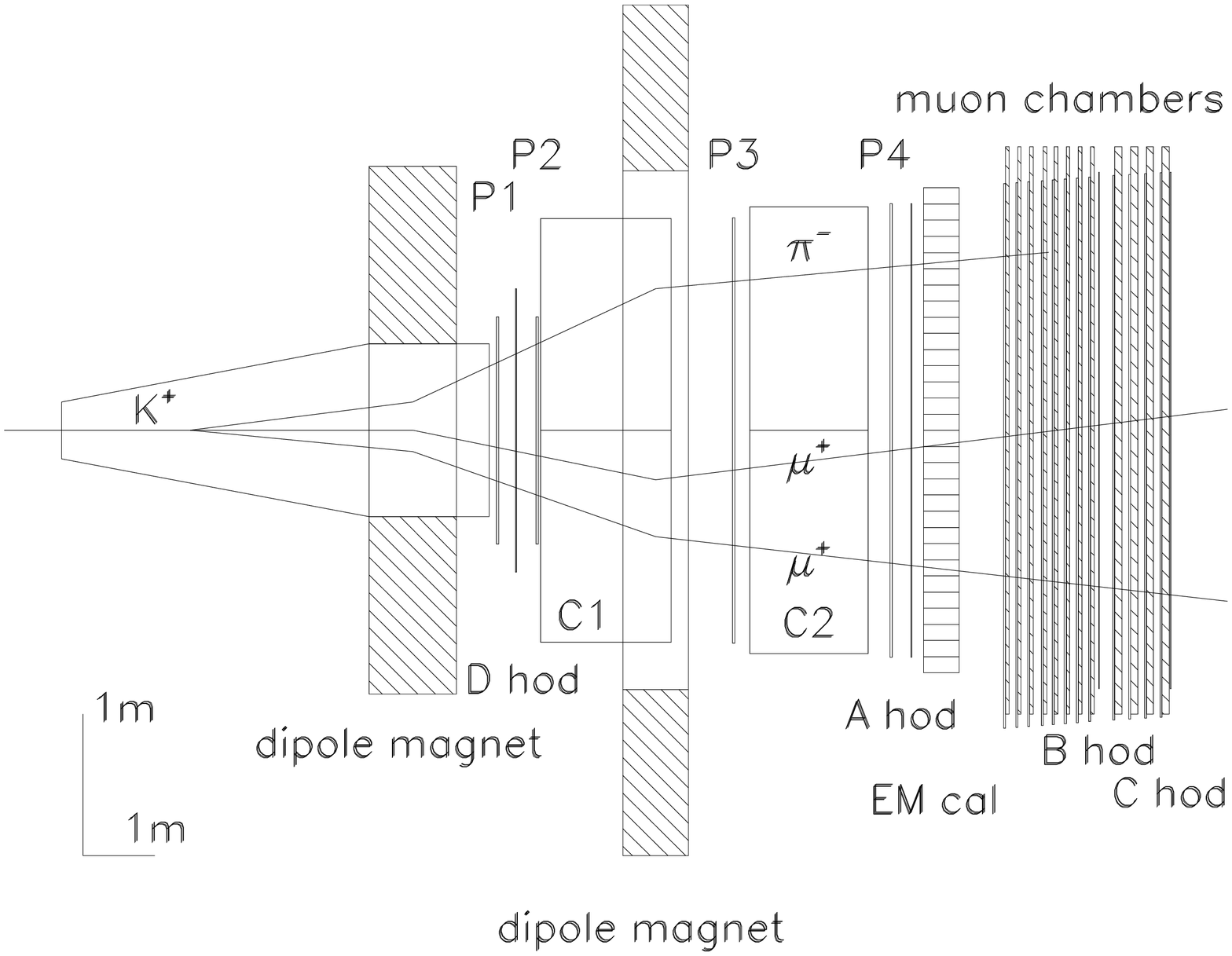,width=80mm}\centering
\caption[Plan view of the E865 detector]
{Plan view of the E865 detector. 
A \mmp\, event is superimposed. 
}
\label{detector}
\end{figure}

The detector (Fig.~\ref{detector}) 
and its performance has been described in other 
publications \cite{e865pme,e865piee,e865pimm,e865nim}.  
The apparatus resided in an unseparated 6 GeV beam 
directly downstream of a 5m 
long evacuated decay volume.  The charged particles from $K^+$ decays 
were first separated in charge by a dipole magnet, then momentum
analyzed in 
a spectrometer system consisting of proportional chambers(P1-P4) 
and another dipole 
magnet.  Particle identification was achieved by two sets of \v{C}erenkov 
counters (C1) upstream and (C2) downstream of the spectrometer magnet, and a 
shashlyk-style electromagnetic
calorimeter downstream of the spectrometer system, followed by a muon 
range stack consisting of steel plates interleaved with
proportional tubes.  For this study,
the \v{C}erenkov counters were filled with methane gas for high $e^\pm$ 
identification efficiency.

The trigger hodoscopes were located directly downstream of the first 
proportional chamber P1 (D-hod), upstream of the calorimeter (A-hod),
and in the middle (B-hod) 
and at the end of the muon stack (C-hod).  The first level trigger was
constructed by requiring
two charged particles on the right, and one charged particle on the left
in 
the A and D hodoscopes and the corresponding calorimeter modules. 
In the next trigger level, particle identification information was
applied.

The trigger designed for \efour\, accepted events 
with $e^+$ but not accompanied by an
$e^-$.  \v{C}erenkov light signals were 
required on the right side of both C1 and C2, and  
both \v{C}erenkov counters on the left were required 
to have signal below one photo-electron,  
to suppress events with an $e^-$ from the 
$\pi^0\rightarrow e^+e^-\gamma$  decay (Dalitz).  

The trigger designed for  \pmm\, decay required
one muon on each side of the detector.  Each muon, for trigger purposes, 
was identified as a spatially
correlated  coincidence between the B and C hodoscope hits. 

In the off-line reconstruction, events are required to have three
charged
tracks from a common decay vertex in the decay volume, a
reconstructed kaon 
momentum consistent with the beam phase space distribution, and a timing 
spread between the tracks consistent with the resolution, typically 
about 0.5ns.     Similar to the 
analysis of the \pmm\, events\cite{e865pimm}, a joint likelihood function
is  constructed  based on the vertex quality, the kaon momentum vector, and 
the track $\chi^2$.  This is used to select events with high kinematic 
quality. 

For \mmp\, events, muons are  required to have momenta greater than 
1.3 GeV/c,  go through the muon stack 
and have corresponding hits in B-hod and C-hod.  
There should be sufficient muon chamber hits associated 
with the track, and energy deposition in the shower calorimeter
should be consistent with minimum ionizing particles.    The trigger
requirement that there be one muon on the left and one on the right is
not
efficient for this decay because  positively charged particles tend to 
populate the right side of the detector.  The majority of the \mmp\, events 
which would be accepted by the trigger would 
have two $\mu^+$'s on the right side 
of the spectrometer system, and one of the $\mu^+$'s crossing to the
left
in the muon system downstream of the calorimeter (see 
Fig.~\ref{detector}).  In a smaller fraction
of events, one of the $\mu^+$'s stays on the left side throughout the
detector, 
and the other 
$\mu^+$ and the pion on the right. Monte Carlo simulation shows that the 
trigger acceptance
for \mmp\, is a factor of 2.7 smaller than that for \pmm.

The background for \mmp\, comes from \ktau(\taus), with both $\pi^+$'s
misidentified as $\mu^+$'s.
Although most of these background events
have the reconstructed $\mu\mu\pi$ mass much lower than $M_K$ because of
the 
mass difference between muon and pion,  events with pion decays in the 
spectrometer magnet can result in $\mu\mu\pi$ mass in the signal
region.  Because those events tend to have worse kinematic characteristics, a
tight cut  on the joint likelihood helps to reduce background. 

\begin{figure}[htb]
\epsfig{figure=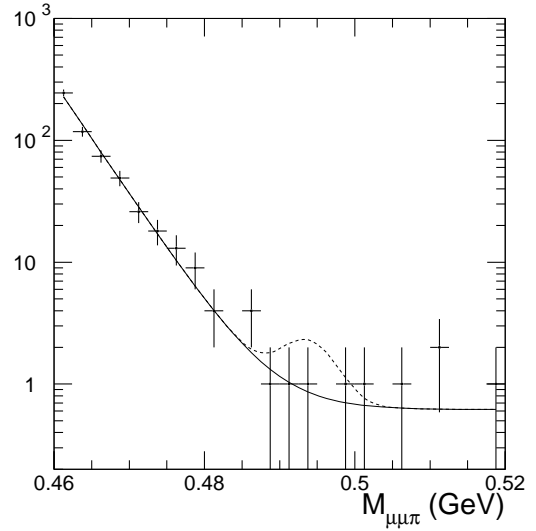,width=80mm}\centering
\caption[pmm2fit]
{
The $\mu\mu\pi$ invariant mass distribution for  \mmp\, candidates.
The points with error bars are data, the solid line is a fit to an
empirical 
function\cite{e865pimm}, 
and the dashed line is a fit that includes a signal at the 
90\% C.L. upper limit. 
}
\label{mmpfit}
\end{figure}

Figure \ref{mmpfit} shows the reconstructed $\mu\mu\pi$ mass 
distribution after a cut on the joint likelihood function.  
The background within the signal region 
($ 0.4875$ GeV $< M_{\mu\mu\pi} < 0.5025$ GeV) 
is estimated by fitting the spectrum with an empirical function
used in the \pmm\ analysis\cite{e865pimm} 
with the signal region excluded. 
There are 5 events in the signal region
where 5.3 background events are expected.
Using the frequentist approach\cite{feldman98}, 
the upper limit on the number of signal
events is 4.8 at the 90\% C.L..
Normalizing to \taus,  we obtain  
an upper limit on the \mmp\ branching ratio:
\begin{eqnarray}
Br(K^+\rightarrow \mu^+\mu^+\pi^-)<3.0\times10^{-9} (90\% \rm{C.L.}).\label{brmmp}
\end{eqnarray}

For $K^+\rightarrow e^+ \pi^\pm \mu^\mp$  events, 
an $e^+$ is required  on the right side with
\v{C}erenkov light associated with the track in both C1 and C2, and 
an E/p ratio of at least 0.8.
The charged pion is required to have no significant signals in the
\v{C}erenkov  counters associated with the track, and calorimeter 
responses consistent with minimum ionizing particles or hadronic
showers.
The $\mu^-$($\mu^+$) is required to be on the  left(right), 
to  reach the B-hod, and to have a range in the muon stack
consistent with its momentum. 
 The minimum momentum for the muon is 0.75 GeV/c.  

The main sources of background for 
$K^+\rightarrow e^+ \pi^\pm \mu^\mp$  decays
are \efour,
when one of the charged 
pions is misidentified as muon,
and \taus, when one $\pi^+$ 
 is mistaken for a muon and the other $\pi^+$ 
misidentified as $e^+$.
The probability of misidentifying a $\pi^+$ as $\mu^+$ 
is 5\% due to pion decays and punchthrough. 
The probability of misidentifying a $\pi^+$ as an $e^+$ is
1.0$\times10^{-4}$. 
This happens when 
the pion deposits most of its energy in the calorimeter, and at the same
time 
there are photoelectrons associated with the track, either 
originating from scintillation or random activity.  
Since the threshold of the \v{C}erenkov counters is 3.5 GeV for muons, 
the high energy muons in the beam halo can produce \v{C}erenkov light.
To reduce this 
misidentification probability, events with additional tracks on the
right side, either
electrons or high energy muons, are rejected from this sample.  

\begin{figure}[htb]
\epsfig{figure=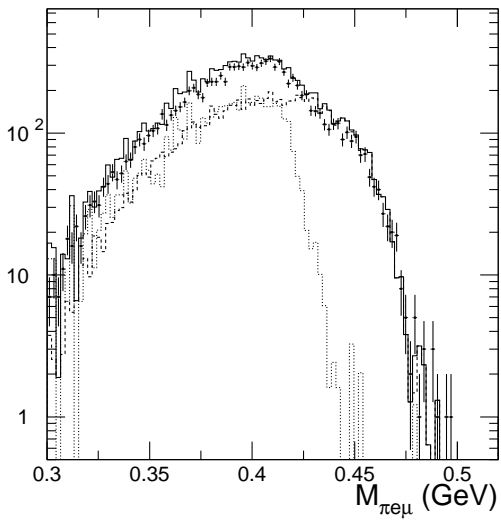,width=80mm}\centering
\caption[empbck]
{
$\pi e \mu  $ invariant mass distribution for \piem\, candidates before a
tight cut on the joint likelihood function. 
The points are data,  the dashed histogram is the \efour\, Monte Carlo 
simulation,  the dotted histogram is the \taus\, Monte Carlo
simulation, 
and the solid histogram is the sum. 
}
\label{empbck}
\end{figure}

In Fig.~\ref{empbck}, data are compared to the Monte Carlo simulation
of 
the background events from \efour\, and \taus, before a tight cut on the 
joint likelihood function is imposed.  {As can be seen}, 
these two decay modes successfully account for the observed background. 

Because of the  undetected 
neutrino in \efour , the \efour\, background is greatly reduced by requiring 
the candidates to have 
a reconstructed kaon momentum vector within the beam phase space.  Due to the large 
difference in rest masses, the \taus\,
background has a lower reconstructed invariant mass.

\begin{figure}[htb]
\epsfig{figure=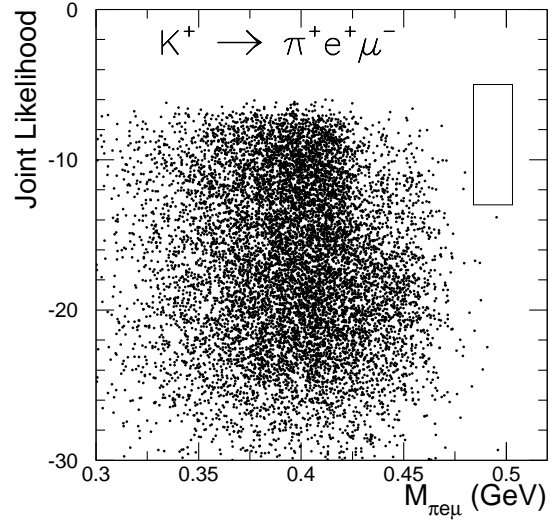,width=80mm}\centering
\caption[empscat]
{ Scatter plot of $M_{\pi e \mu}$ and joint likelihood function
for \piem\, candidate events.  The box indicates the signal region.
}
\label{empscat}
\end{figure}

\vspace{-0.8cm}
\begin{figure}[htb]
\epsfig{figure=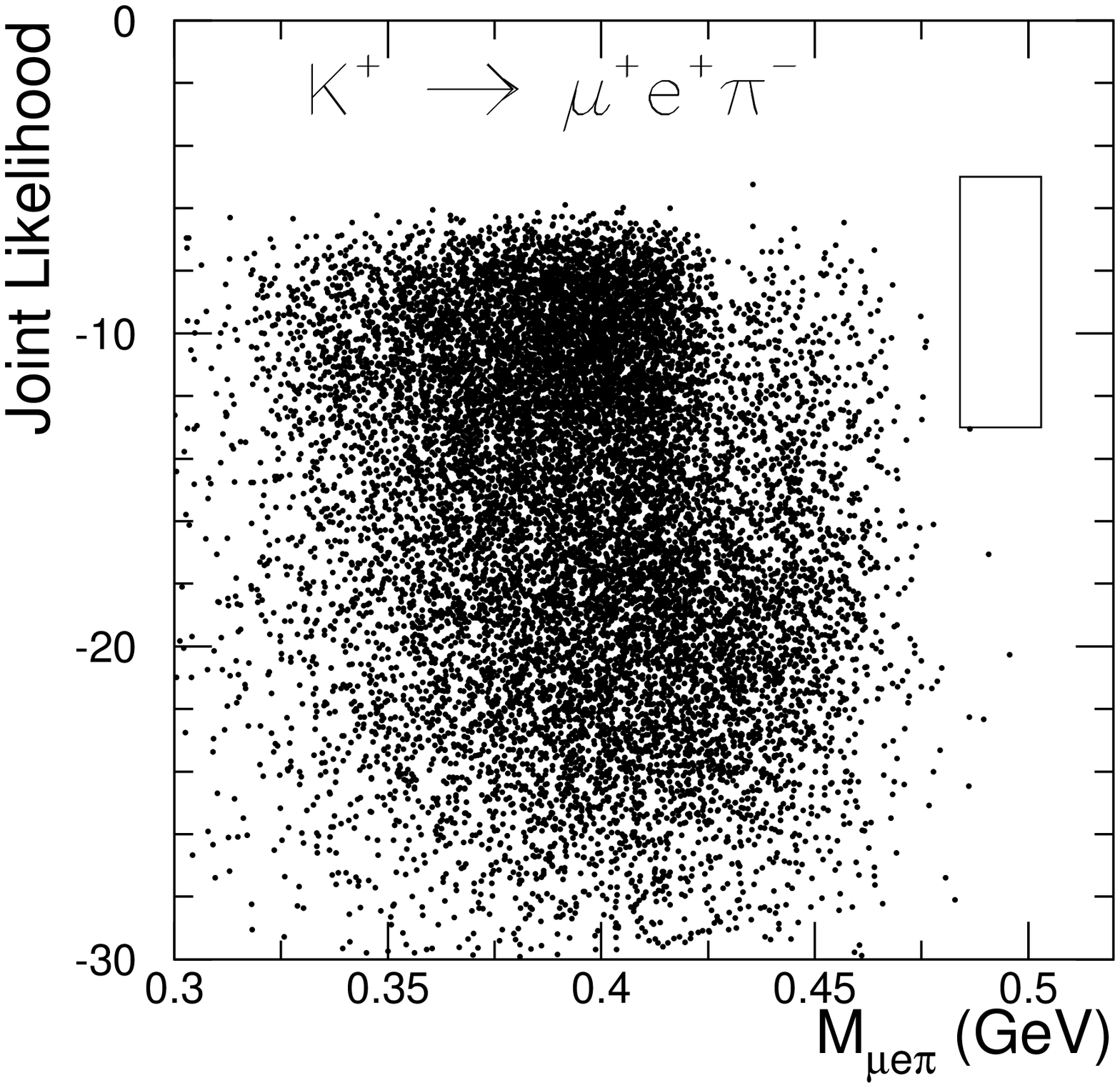,width=80mm}\centering
\caption[epmscat]
{ Similar to Fig.~\ref{empscat}, but for \mep.
}
\label{epmscat}
\end{figure}

Another potential background is \kpp, followed by $\pi^0$ Dalitz decay. 
Since 
the misidentification probability of $e^-$ as $\mu^-$ is negligible, it
does 
not contribute to \piem\ background. For \mep\, events, the reduction of the 
remaining Dalitz background to a negligible level 
is achieved by requiring $M_{ee}>50$MeV, where $M_{ee}$ is the 
invariant mass of $e^+$ and $\pi^-$ with the $\pi^-$ mass assigned to be 
the electron mass. 

Figures \ref{empscat} and \ref{epmscat} are the scatter plots of
the invariant mass of the reconstructed candidate events vs. 
the joint likelihood function.  The boxes indicate the signal region, 
which covers $\pm 3\sigma$ in mass, and 80\% acceptance in joint
likelihood function.
No signal events are observed. 

\begin{figure}[htb]
\epsfig{figure=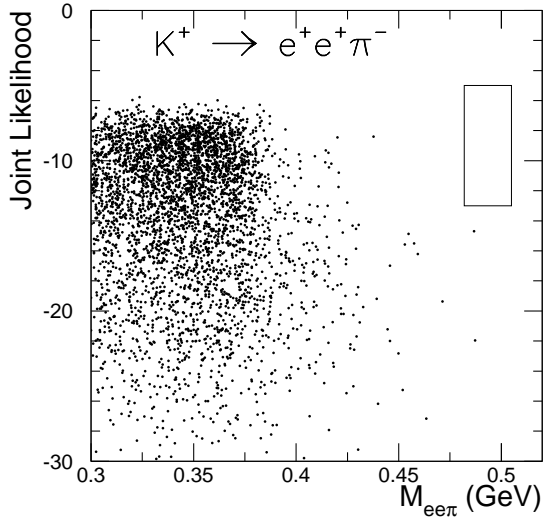,width=80mm}\centering
\caption[eepscat]
{
 Similar to Fig.~\ref{empscat}, but for \eep.
}
\label{eepscat}
\end{figure}

The search for \eep\ applies particle identification conditions
as described above.  The background comes from \efour\, where the $\pi^+$ is
misidentified 
as an $e^+$, and from \taus\, where both $\pi^+$'s are misidentified as
$e^+$'s.
Because of the more 
significant mass difference between $\pi$'s and $e$'s these 
backgrounds are far away from the \eep\ signal region. 
Figure \ref{eepscat} shows the scatter plot of $M_{eep}$ vs. the joint 
likelihood function.   Again, there are no events in the signal box. 
The background events in this plot are correctly accounted for by \taus\,
and \efour.

The observation of  no signal event for \kpem\,  implies also a null 
result for the search of \pizem(\pzem), through the decay
\kpitwo(\pitwo).

Table \ref{accept} lists the acceptances for the decays of interest. 
Normalized to \efour\,
decay, 
the null results of these searches are expressed in term of the 90\%
C.L.~upper 
limit of the branching ratios, 

\begin{table}[htb]
\begin{center} 
 \begin{tabular}{|c|c|}
Decay               & Acceptance (\%) \\ \hline
$\pi^+\pi^-e^+\nu$  &        3.93     \\ 
$ e^+e^+\pi^-$      &        1.54     \\ 
$\pi^+e^+\mu^-$     &        1.90     \\ 
$\mu^+e^+\pi^-$     &        1.97     \\ 
$ \pi^+\pi^0, \pi^0\rightarrow e^+\mu^-$      &        1.38     \\ 
\hline  
$\pi^+\pi^+\pi^-$   &        6.25     \\ 
$\mu^+\mu^+\pi^-$   &        0.71     \\ 
\end{tabular}  
\end{center} 
\caption[]
{ 
 The acceptances for \efour, \mep, \eep, \piem, \taus, \mmp\, and \pitwo+\pzem\, 
decays,  using Monte Carlo 
 simulation.  Detector efficiencies and effects of all kinematic and
particle  identification cuts are  included.  
}
\label{accept}
\end{table} 

\begin{eqnarray}
 Br_s < \frac{N_s\times Br_{Ke4}\times Acc_{Ke4}}{N_{Ke4}\times Acc_s}
\nonumber \\
Br(K^+\rightarrow  e^+e^+\pi^- )  <  6.4\times10^{-10} \\
Br(K^+\rightarrow \pi^+e^+\mu^-)  <  5.2\times10^{-10} \\
Br(K^+\rightarrow \mu^+e^+\pi^- ) <  5.0\times10^{-10} \\
Br(\pi^0\rightarrow e^+\mu^-)     <  3.4\times10^{-9} 
\end{eqnarray}
$Acc_{Ke4}$($Acc_s$) is the acceptance to \efour\ (signal) decay, and 
$N_s=2.44$, $Br_{Ke4}=3.91\times10^{-5}$, $N_{Ke4}=378,000.$ 
For \pzem, the \pitwo\, branching ratio of 0.21 is taken into account. 

The limits on \piem, \mep\, and \eep\, represent an improvement of more 
than a factor  of 10 over the previous searches\cite{diamant}.  
The upper limit on Br(\pizem) and 
our result of Br($\pi^0 \rightarrow \mu^+e^-)<3.8\times 10^{-10}$ \cite{e865pme}
can be compared 
to the previous best limit of 
$[Br(\pi^0\rightarrow\mu^+ e^-)+Br(\pi^0\rightarrow\mu^- e^+)]
<1.72\times10^{-8}$\cite{e799}. 
The new upper limit on \mmp (Eq.~\ref{brmmp}) is a factor of 50000 
better than  the previous experimental bound. 
The implications of this result are discussed in 
 \cite{litt00}. 

We thank L.~Littenberg, R.~Shrock and K.~Zuber for useful discussions.
We gratefully acknowledge the contributions to the success of
this experiment by 
the staff and management of the AGS at the Brookhaven National
Laboratory, and the technical staffs of the participating institutions.
This work was supported in part by the U. S. Department of Energy, 
the National Science Foundations of the USA(REU program), 
Russia and Switzerland, and
the Research Corporation.

\end{document}